\documentclass{rspublic}

\begin{document}

\title{Filtered Lattice Boltzmann Collision Formulation Enforcing Isotropy and Galilean Invariance}

\author{Hudong Chen, Raoyang Zhang, Pradeep Gopalakrishnan}
\affiliation{Dassault Systemes, 185 Wyman Street, Waltham, MA 02451-1223}
\date{\today}
\maketitle






The central framework of a filtered lattice Boltzmann collision operator
formulation is to remove hydrodynamic moments that are not supported by the 
order of isotropy of a given lattice velocity set. Due to the natural moment 
orthogonality of the Hermite polynomials, the form of a filtered collision
operator is obtained directly via truncation of the Hermite expansion. 
In this paper, we present an extension of the filtered collision operator 
formulation to enforce Galilean invariance.  This is accomplished by 
representing hydrodynamic moments in the relative reference frame with
respect to local fluid velocity. The resulting collision operator 
has a compact and fully Galilean invariant form, and it can then be
exactly expressed in terms of an infinite Hermite expansion. 
Giving a lattice velocity set
of specific order of isotropy, a proper truncation of this expansion 
can be directly determined. Higher order terms are retained in the truncation 
if a higher order lattice velocity set is used, so that Galilean invariance is 
attained asymptotically. The previously known filtered collision operator forms 
can be seen as a limit of zero fluid velocity.

\newpage

\section{Introduction}

Among the main themes of the program “Turbulent Mixing and Beyond” (TMB),
and in particular the 10th Anniversary Program TMB-17, is the analysis of 
theoretical principles behind modeling of non-equilibrium dynamics on the 
continuum as well as kinetic scales. Here we attempt at a universal approach 
to description of a wide class of kinetic systems that possess a well developed 
hydrodynamic limit. While von Neumann alluded to theory of non-equilibrium systems
as to theory of non-elephants, we believe that the TMB program is not at 
all dismissive of, but is actually quite in line with the giant. Insofar as 
one may speculate that von Neumann implied that physical systems away from 
equilibrium are less universal, this needs not necessarily mean that universality 
is totally absent. Instead, systematic effort will be required to study common 
features of non-equilibrium phenomena, and that is the focus of TMB program 
(and of course much other work over the last several decades). Among many 
aspects of the interplay between non-equilibrium and equilibrium is a 
question of just how much information is lost in the process of constructing 
effective macroscopic description out of non-equilibrium that is present at 
smaller scale. In this work we systematically derive non-equilibrium distribution 
functions in a general way that respects system’s symmetries while preserving 
the moments of distribution function responsible for conserved hydrodynamic properties, 
potentially neglecting the information pertaining to non-equilibrium properties 
that are not relevant for adequate large scale description. This general 
method might be useful for trying to understand commonalities between 
non-equilibrium phenomena that the TMB program aspires to promote.

Lattice Boltzmann Methods (LBM) have been developed as a useful method for 
analyzing fundamental and computational aspects of fluid dynamics during past few decades \cite{Benzi,ChenDoolen}. 
A lattice Boltzmann system is a mathematical model in that its 
underlying dynamics resides upon the fundamental physics of kinetic theory. Namely
it involves motion of many particles according to the framework
of Boltzmann equation \cite{Huang,Cerci}.
On the other hand, unlike the classical kinetic theory that is defined on continuous 
phase-space and time, the latter in a lattice Boltzmann system are all discrete.
Specifically, in a lattice Boltzmann model, particles are only allowed 
to have a handful set of values, and they correspond to the links from
one site to its near neighboring sites on a Cartesian Bravais lattice (cf. \cite{FHP1}). 
Therefore, at each constant time increment the particles hop from one lattice site
to their respective neighboring sites according to their own velocity values. 
The dynamics of a lattice Boltzmann system is governed by a  
lattice Boltzmann equation \cite{Chen91,Chen92,Qian92}.
Using the so called ``lattice units'' so that $\Delta x = \Delta t = 1$,
a lattice Boltzmann equation is commonly expressed in a generic difference-form below,
\begin{equation}
\label{lbe}
f_i({\bf x} + {\bf c}_i, t + 1) - f_i({\bf x},t) = C_i({\bf x},t)
\end{equation}
where $f_i({\bf x},t)$ is the particle distribution function denoting
number of particles with a velocity value ${\bf c}_i$ at a lattice site
${\bf x}$ and time $t$.  All possibly allowed velocity values
are contained in a pre-assigned set of values $\{ {\bf c}_j \; ; \; j = 0, \ldots , b \}$.  
Here $b$ is a constant integer ($b \sim 20$). Different lattice Boltzmann
models may have different sets of values.  
Conceptually, the left-hand side of eqn.(\ref{lbe}) represents advection of particles 
from one lattice site to another from time $t$ to $t + 1$. Since every particle velocity 
value is a link between two lattice sites, both ${\bf x}$ and ${\bf x} + {\bf c}_i$ are 
sites on the lattice.  The collision process is denoted by the term $C_i({\bf x},t)$
on the right-hand side of eqn.(\ref{lbe}).

The discrete nature of a lattice Boltzmann model present a simplistic abstraction of a 
realistic particle system, so that some fundamental physics may be revealed more
clearly. This is closely analogous in spirit to the Ising model for realistic 
ferromagnetic substances. Furthermore, due to its kinetic theory origin 
as well as such a discrete mathematical representation, 
it serves not only an alternative efficient computation method
for numerically solving the Navier-Stokes fluid equations, but also 
offers a way to study fluid flow physics in a broader fluid regime and  
beyond (cf. \cite{succi} and references therein).  
Indeed, there have been significant successes in LBM in simulation of complex
fluid flows and making a significant impact on real world applications \cite{science}.     

On the other hand, the discrete structure of LBM also poses a fundamental challenge.
Unlike the continuum kinetic theory in which particles can
have all possible velocity values, in LBM there is only a finite set of 
velocity values that are chosen according to
a given Cartesian lattice. Consequently, there is a special frame of reference 
corresponding to the lattice being at rest. This means that the fundamental law 
of Galilean invariance is not going to be exactly satisfied.
How to formulate proper LBM models that recover Galilean invariance at various
approximation levels is at the heart of the LBM research since its very beginning 
\cite{Chen92,Qian92}. However, most of the pre-existing works are
focused on constructing appropriate equilibrium distributions and 
corresponding discrete velocity sets \cite{Molvig,Shan06,Chen-mom,ChenGoldOrszag}.
The work presented here is pertaining to the non-equilibrium
distribution of particles in the overall LBM theoretical formulation 
together with its associated proper collision process.
     
The non-equilibrium property is intrinsically associated with
the collision process in a many-particle kinetic
system \cite{Cerci}. As in a realistic particle system, the collision process is one of 
the two fundamental micro-dynamical processes in LBM, 
- the other one is the advection process. A collision process is
essential for individual particles to interact and form a collective fluid-like
behavior. During a collision process, particles exchange mechanical, thermal and
other physical properties subject to conservation laws. 

From the distribution function, one can construct hydrodynamic moments of $n$-th order,
\begin{equation}
\label{N-moments}
{\bf M}^{(n)}({\bf x},t) \equiv 
\sum_i \underbrace{{\bf c}_i \ldots {\bf c}_i}_\text{n} f_i({\bf x},t) 
\end{equation}
Among these the most fundamental ones are the first two moments 
corresponding to local mass and momentum, namely,
\begin{eqnarray}
\label{mme}
\rho({\bf x},t) &\equiv& {\bf M}^{(0)}({\bf x},t) = \sum_i f_i({\bf x},t) \nonumber \\
\rho({\bf x},t) {\bf u}({\bf x},t) &\equiv& {\bf M}^{(1)}({\bf x},t)
= \sum_i {\bf c}_i f_i({\bf x},t) 
\end{eqnarray}
For simplicity and without loss of generality of the basic theoretical framework, 
here we discuss the so called ``isothermal'' 
LBM models in which the energy conservation is not enforced.

Due to conservation laws, the first two moments are invariant under collisions, so that
\begin{equation}
\label{inv}
\sum_i \chi_i f_i({\bf x},t) = \sum_i \chi_i f'_i({\bf x},t) = \sum_i \chi_i f^{eq}_i({\bf x},t) 
\end{equation}
where $\chi_i = 1$ or ${\bf c}_i$. In the above, 
$f'_i({\bf x},t)$ and $f^{eq}_i({\bf x},t)$ denote the so called ``post-collide''
and equilibrium distribution functions, respectively, 
\begin{equation}
\label{post}
f'_i({\bf x},t) \equiv f_i({\bf x},t) + C_i({\bf x},t) 
\end{equation}
Hence (\ref{inv}) apparently leads to the vanishing values of 
the collision operator for the two conserved hydrodynamic moments, 
\begin{equation}
\label{conserv}
\sum_i \chi_i C_i({\bf x},t) = 0 
\end{equation}
Besides the two conserved moments, another essential
hydrodynamic moment for an isothermal fluid system is 
the so called momentum flux tensor,
\begin{equation}
\label{Mflux}
{\bf M}^{(2)} ({\bf x},t) \equiv \sum_i {\bf c}_i{\bf c}_i f_i({\bf x},t)
= {\bf \Pi}^{eq} ({\bf x},t) + {\bf \Pi}^{neq} ({\bf x},t) 
\end{equation}
where
\begin{eqnarray}
\label{Mflux-eqn}
{\bf \Pi}^{eq} ({\bf x},t) &\equiv& \sum_i {\bf c}_i{\bf c}_i f^{eq}_i({\bf x},t) 
\nonumber \\
{\bf \Pi}^{neq} ({\bf x},t) &\equiv& \sum_i {\bf c}_i{\bf c}_i 
(f_i({\bf x},t) - f^{eq}_i({\bf x},t)) 
\end{eqnarray}
corresponding to the equilibrium and non-equilibrium part of momentum flux, respectively.
For an isothermal LBM to obey the fluid physics in a specific regime of interest,
the first three moments (in (\ref{mme}) and (\ref{Mflux})) 
must recover the same hydrodynamic forms as that of the continuum Boltzmann 
kinetic theory \cite{Shan06,Chen-mom}.

The simplest form of a collision operator is the so called 
``BGK'' form \cite{BGK,Chen91,Qian92},
\begin{equation}
\label{BGK}
C_i = - \frac {f_i - f^{eq}_i} {\tau} 
\end{equation}
where $\tau$ is a relaxation time.  It automatically satisfies 
the conservation requirements of (\ref{conserv}) due to
(\ref{inv}). On the other hand, a BGK collision operator
also generates in general moments of all orders 
besides the momentum flux tensor. The higher order moments 
are not all supported by a given lattice with a finite degree of isotropy,
so that they manifest as unphysical numerical artifacts, similar to the aliases in
the standard spectral methods. Hence a filtered collision form needs to be formulated \cite{Zhang,Rot,chopard}. Unfortunately these pre-existing filtered forms are
only valid for fluid flows at a very low fluid velocity 
with respect to a lattice at rest.
The new work presented here is on how an appropriate filtered 
collision form should be theoretically constructed 
so that Galilean invariance is properly approximated for a lattice
with a given finite degree of isotropy as well as how such
an invariance is obeyed asymptotically in the limit of infinite
lattice isotropy for the non-equilibrium distribution.    

\section{Basic Filtered Collision Formulation}

Formulation of a filtered collision process has been realized via 
a regularization procedure \cite{Rot,Zhang,chopard}.
First of all, when a lattice velocity set is specified, 
then its order of isotropy is determined, as described by the subsequent
condition. Namely, for a given set of lattice velocity vectors 
$\{ {\bf c}_i \; ; \; i = 0, 1, \ldots , b \}$, it
is $2N$-th order isotropic if the set satisfies
up to $2N$ \cite{Chen-mom,ChenGoldOrszag} the following properties,
\begin{eqnarray} 
E^{(n)} \equiv \sum_i w_i \underbrace{{\bf c}_i \ldots {\bf c}_i}_\text{n} 
= \begin{cases}                                                       
T_0^{\frac{n}{2}} \Delta^n, & n = 2, 4,...., 2N \label{eq.identity} \\
							0, & n = \text{odd integers}
                                                      \end{cases}
\label{iso}
\end{eqnarray}
where $\Delta^{n}$ is the $n$-th order Kronecker delta function tensor 
(see \cite{Chen-mom,ChenGoldOrszag}), $w_i$ is a constant weighting factor. 
$T_0$ is a constant (for instance $T_0 = 1/3$ in lattice units 
for D3Q15 and D3Q19 lattices). 
Obviously, $N$ is always a finite integer for any lattice velocity set 
comprised of a finite number of velocity values.
In particular, the so called D2Q9, D3Q15 and D3Q19 lattices 
have $2N = 4$ \cite{Qian92}. On the other hand, lattices such as D3Q39 admits
$2N = 6$ or higher \cite{Shan06}.  Consequently, in order for the hydrodynamic 
moments (\ref{N-moments}) to correspond to that of a physical
fluid system, it must be fully expressible in terms of the tensors 
that are supported by a lattice with sufficient order of isotropy 
according to the condition (\ref{iso}) \cite{Shan06,Chen-mom}. This is easily 
understood when expressing the distribution function in (\ref{N-moments})
in terms of Hermite polynomials. For instance,
for the equilibrium distribution function, we have \cite{Shan06}
\begin{equation}
\label{hermite}
f^{eq}_i({\bf x}, t) = w_i \rho({\bf x}, t) \sum_{n=0}^\infty
\frac{H^{(n)}(\xi_i)}{n!} {\bf V}({\bf x}, t)^{[n]}
\end{equation}
where ${\bf V}({\bf x}, t)^{[n]}$ is a short notation for a 
direct product of vector ${\bf V}({\bf x}, t)$ n-number of times.  
${\bf V}({\bf x},t) = {\bf u}({\bf x},t)/T_0^{1/2}$ 
and $\xi_i = {\bf c}_i/T_0^{1/2}$.  The n-th order Hermite
function $H^{(n)}(\xi_i)$ is a n-th rank tensor generalization of 
the standard (scalar) n-th order Hermite polynomial \cite{Grad}. 
It can be straightforwardly shown that $H^{(n)}(\xi_i)$ is a linear
combination of direct product of ${\bf c}_i^{[m]}$ ($\{ m = 0, \ldots , n\}$).
Thus, retaining only the supported tensor
terms is equivalent to a specific truncation of the Hermite
expansion.  In particular, in order to ensure a physically correct equilibrium 
momentum flux, the equilibrium distribution function should only 
contain Hermite polynomials up to at most $2N -1$ \cite{Chen-mom,ChenGoldOrszag,Molvig}.
Terms in (\ref{hermite}) higher than $2N - 1$ contain discrete artifacts.

Likewise, the essential idea of the filtered collision operator formulation
is to apply the above procedure on the non-equilibrium part of the distribution
function \cite{Zhang}.
Together with the BGK operator, we can express the post-collide distribution (\ref{post}) as
\begin{equation}
\label{post-1}
f'_i({\bf x},t) \equiv f^{eq}_i({\bf x},t) + (1 - \frac {1} {\tau}) f^{neq}_i({\bf x},t) 
\end{equation}
with
\begin{equation}
\label{neq}
f^{neq}_i({\bf x},t) \equiv f_i({\bf x},t) - f^{eq}_i({\bf x},t) 
\end{equation}
Projecting $f^{neq}_i({\bf x},t)$ onto a truncated Hermite basis, it becomes
\begin{equation}
\label{neq-trun}
{\hat f}^{neq}_i({\bf x}, t) = w_i \sum_{n=2}^{2N-1}
\frac{H^{(n)}(\xi_i)}{n!} \frac {{\bf a}^{(n)}}{T_0^{n/2}}
\end{equation}
Notice the summation starts at $n = 2$ due to vanishing of the first
two moments (see (\ref{conserv})).  The coefficient ${\bf a}^{(n)}$ above is given by,
\begin{equation}
\label{coeff}
{\bf a}^{(n)}({\bf x}, t) 
= \sum_i (f_i({\bf x},t) - f^{eq}_i({\bf x},t)) H^{(n)}(\xi_i)T_0^{\frac{n}{2}}
\end{equation}
Specifically, 
\begin{equation}
\label{coeff-2}
{\bf a}^{(2)}({\bf x}, t) \equiv {\bf \Pi}^{neq}({\bf x},t)
= \sum_i {\bf c}_i {\bf c}_i (f_i({\bf x},t) - f^{eq}_i({\bf x},t)) 
\end{equation}
\begin{equation}
\label{coeff-3}
{\bf a}^{(3)}({\bf x}, t) \equiv {\bf Q}^{neq}({\bf x},t)
= \sum_i {\bf c}_i {\bf c}_i {\bf c}_i (f_i({\bf x},t) - f^{eq}_i({\bf x},t)) 
\end{equation}
Due to the orthogonality of Hermite polynomials,
the filtered distribution ${\hat f}^{neq}_i({\bf x}, t)$
gives identical values as the original $f^{neq}_i({\bf x},t)$
for moments up to $2N + 1$, while all the higher moments vanish.  This
is the essence of the filtered collision formulation.
For the most commonly used lattices with the order of isotropy $2N = 4$, 
one obtains an explicit form of (\ref{neq-trun}) below \cite{Zhang,Rot,chopard}
\begin{eqnarray}
\label{filter2}
{\hat f}^{neq}_i({\bf x}, t)
&=& \frac{w_i}{2T_0} [\frac{{\bf c}_i {\bf c}_i} {T_0} - {\bf I}] : 
{\bf \Pi}^{neq}({\bf x},t) \nonumber \\
&+& \theta \frac{w_i}{6T_0^2} [\frac{{\bf c}_i {\bf
    c}_i {\bf c}_i}{T_0} - 3 {\bf c}_i  {\bf I}] \vdots {\bf Q}^{neq}({\bf x},t)
\end{eqnarray}
where ${\bf I}$ is the 2nd rank unity tensor. Parameter $\theta$ in the above 
can be either zero or other scalar values to account for a 
non-unity Prandtl number \cite{ShanPrandtl,Prandtl}.
With (\ref{filter2}), the regularized lattice Boltzmann equation
with the filtered collision process is given by,
\begin{equation}
\label{lbe-reg}
f_i({\bf x} + {\bf c}_i, t + 1) = f^{eq}_i({\bf x},t) 
+ (1 - \frac {1} {\tau}) {\hat f}^{neq}_i({\bf x},t) 
\end{equation}

\section{Recovering Galilean Invariance in Filtered Collision Formulation}

It can become clear that the filtered collision forms above
are only for situations of small fluid flow velocity. 
From the basic principle of Galilean invariance, the distribution functions 
(equilibrium and non-equilibrium) of a physical many-particle system 
should be functions of particle velocities relative to their local fluid velocity
rather than any particular reference frame (for instance, the reference frame
of the lattice at rest).  
Indeed, it can be shown that the infinite Hermite expanded equilibrium 
distribution function (\ref{hermite}) admits a compact and
Galilean invariant form similar to the Maxwell-Boltzmann distribution \cite{Molvig,Chen-mom}, 
\begin{equation}
\label{feq_full}
f^{eq}_i({\bf x}, t) = w_i e^{e_i/T_0} \rho({\bf x},t) 
e^{- {\bf c}'_i({\bf x},t)^2/2T_0}
\end{equation}
where ${\bf c}'_i({\bf x}, t) \equiv {\bf c}_i - {\bf u}({\bf x}, t)$
is the particle velocity relative to the local fluid velocity.  
(Clearly ${\bf c}_i$ is the particle velocity in the reference frame of the lattice at rest.) 
$e_i \equiv {\bf c}^2_i/2$. 
Therefore, we should look for a similar compact Galilean invariant form for
the non-equilibrium distribution function. 

The first step for accomplishing the above is to redefine the proper hydrodynamic
moments. Instead of (\ref{coeff-2}) and (\ref{coeff-3}) that are defined
in the particular absolute reference frame of lattice at rest, 
they need be replaced by the corresponding ones in the relative reference frame 
in terms of the relative particle velocities,
\begin{equation}
\label{Pi_def}
{\bf \Pi}^{neq}({\bf x},t) = \sum_i {\bf c}'_i({\bf x}, t) {\bf c}'_i({\bf x},t)
   [f_i({\bf x}, t) - f^{eq}_i({\bf x}, t)]  
\end{equation}
\begin{equation}
\label{Q_def}
{\tilde {\bf Q}}^{neq}({\bf x}, t) = \sum_i {\bf c}'_i({\bf x}, t) {\bf c}'_i({\bf x},t) {\bf
 c}'_i({\bf x}, t) [f_i({\bf x}, t) - f^{eq}_i({\bf x}, t)]
\end{equation}
Interestingly, due to mass and momentum conservation, the relative
non-equibrium momentum flux in (\ref{Pi_def}) turns out to be identical to (\ref{coeff-2}).  
On the other hand, the relative non-equilibrium energy flux 
${\tilde {\bf Q}}^{neq}({\bf x}, t)$ in (\ref{Q_def}) 
is not the same as ${\bf Q}^{neq}({\bf x},t)$ in (\ref{coeff-3}).
Without loss of generality on the fundamental concept but for clarity, in the rest of the section 
we describe in detail the extended filter formulation to only include 
the relative non-equilibrium momentum flux. The procedure for a 
general formulation involving the non-equilibrium energy flux is 
straightforward and is provided in the Discussion section. 

Now we seek a proper Galilean invariant formulation of non-equilibrium distribution. 
Since the full functional form (\ref{feq_full}) of the equilibrium distribution is Galilean invariant,
the goal is thus to tie the non-equilibrium distribution to its equilibrium counterpart. 
First we briefly discuss the basic concept of how the non-equilibrium distribution can be expressed 
as a function of the equilibrium distribution in the framework of
fundamental physics of kinetic theory (cf. \cite{Huang}). 
The equilibrium and non-equilibrium distributions are intrinsically related to each other 
via the dynamics of the Boltzmann kinetic equation. Hence, the non-equilibrium distribution 
can in principle be expressed in terms of the equilibrium distribution function. 
Indeed, such an explicit functional form can either be 
expressed as an infinite series in powers of spatial and temporal derivatives via 
the so called Chapman-Enskog expansion \cite{CE}, or an exact compact form under some specific
microscopic conditions \cite{COS}. 
Consequently, one sees that the non-equilibrium distribution is simply a result of 
spatial-temporal inhomogeneity of the equilibrium distribution. The latter is 
solely a function of the fundamental conserved fluid quantities $\rho$, ${\bf u}$ and $T$.  
Hence the inhomogeneity of the equilibrium distribution can be expressed in terms of 
gradients of these fundamental conserved fluid quantities and their higher order derivatives. 
But due to the special Maxwell-Boltzmann exponential form of the equilibrium distribution
function and vector-tensor symmetries, one can realize that the fundamental form
of the non-equilibrium function is proportional to the equilibrium distribution
multiplied by some proper scalar contractions of vector and tensors 
corresponding to gradients and the higher derivatives
of the conserved fluid quantities. 
From the fundamental physics point of view, the results above are clear. That is, a fluid
is in equilibrium when it is spatially homogeneous, and generation of non-equilibrium is
due to inhomogeneity in the fluid flows. Furthermore,
the gradients and higher order derivatives of these conserved fluid quantities 
have direct relationships to their corresponding hydrodynamic fluxes \cite{forster}. 
Indeed, in the Navier-Stokes fluid regime such relationships are explicitly defined
via the Newtonian fluid constitutive relations. Namely, the momentum flux tensor is
linearly proportional to the gradient of fluid velocity (i.e., the rate of strain tensor), and 
the energy flux is linearly proportional to the gradient of temperature.
In the reference frame of local fluid flow velocity,
there are only two such hydrodynamic fluxes, namely the non-equilibrium momentum 
flux ${\bf \Pi}^{neq}$ and the non-equilibrium energy 
flux ${\bf Q}^{neq}$, respectively.  
Ignoring the energy flux for isothermal flows, one finally obtains in the continuous kinetic
theory an explicit form for the non-equilibrium distribution 
\begin{equation}
\label{continuum}
f^{neq} \sim f^{eq} * [{\bf c}' {\bf c}' 
- \frac{1}{D} {\bf c}'^2 {\bf I}] : {\bf \Pi}^{neq}
\end{equation}
where ${\bf c}' = {\bf c} - {\bf u}$
is the particle velocity relative to the local mean flow in
the continuum kinetic theory. The overall meaning of the relevant physics is 
far beyond what is discussed here \cite{Cerci}. Inspired by this concept, we seek in LBM 
a similar functional expression for the non-equilibrium distribution.  
Rewriting (\ref{filter2}) and dropping off the terms proportional to ${\bf Q}^{neq}$, 
it is simply \cite{Zhang,Rot,chopard}
\begin{equation}
\label{filter-Pi}
{\hat f}^{neq}_i({\bf x}, t)
= \frac{w_i}{2T_0} [\frac{{\bf c}_i {\bf c}_i} {T_0} - {\bf I}] : {\bf \Pi}^{neq}({\bf x},t) 
\end{equation}
Comparing this form to that of (\ref{continuum}) and observing that 
$w_i\rho({\bf x}, t)$ is exactly the equilibrium distribution (\ref{feq_full}) 
at zero fluid velocity (${\bf u}({\bf x}, t) = 0$), 
we immediately recognize that (\ref{filter-Pi}) is the non-equilibrium distribution 
in the limit of zero local fluid velocity.  Therefore, to extend it 
for finite fluid velocity, one can simply 
replace $w_i$ by $f^{eq}_i({\bf x}, t)/\rho({\bf x}, t)$ together with 
replacing everywhere the absolute lattice velocity ${\bf c}_i$ by 
its corresponding relative velocity ${\bf c}'_i({\bf x},t)$.  
Hence, we obtain a compact Galilean invariant form for the non-equilibrium
distribution function similar to 
(\ref{continuum}) in the continuum kinetic theory,
\begin{equation}
\label{filter-coll}
{\hat f}^{neq}_i({\bf x}, t) = \frac{f^{eq}_i({\bf x}, t)}{2 \rho({\bf x}, t) T_0} 
[\frac{{\bf c}'_i({\bf x}, t) {\bf c}'_i({\bf x}, t)}{T_0} 
- {\bf I}] : {\bf \Pi}^{neq}({\bf x}, t)
\end{equation}
where ${\bf \Pi}^{neq}({\bf x}, t)$ is given by (\ref{coeff-2}) or equivalently by (\ref{Pi_def}),
and $f^{eq}_i({\bf x}, t)$ is given by (\ref{feq_full}). 

Obviously, the fully Galilean invariant forms of the equilibrium distribution (\ref{feq_full})
and the non-equilibrium distribution (\ref{filter-coll})
are only realizable if the supporting lattice velocity set has 
an infinite order of isotropy. This is impossible
with any given lattice having a finite set of discrete lattice
velocity values.  Nevertheless, we can achieve the goal of Galilean invariance 
approximately up to a finite order corresponding to the 
isotropy of a given supporting lattice.
The full Galilean invariance is then approached asymptotically as we
choose a lattice with an even higher order of isotropy.
The procedure is the same as the basic filtering formulation described in
the previous section.  That is, we first express (\ref{feq_full}) 
and (\ref{filter-coll}) in terms of infinite Hermite expansions,  
and then we apply a proper truncation on the two expansions according to 
the order of isotropy of a given supporting lattice velocity set.
Due to the orthogonality of the Hermite polynomials, the truncated Hermite
expansions give identical results for moments (\ref{N-moments}) (up to 
the corresponding order supported by the given lattice) as the full
distribution functions (\ref{feq_full}) and (\ref{filter-coll}). On the other
hand, unlike the infinite Hermite expansions,
all the higher moments vanish for the truncated expansions. 

The Hermite expanded form of the equilibrium distribution (\ref{feq_full}) 
is already available in (\ref{hermite}) above.
The compact form of the non-equilibrium distribution function
(\ref{filter-coll}) can also be exactly expressed (see Appendix) 
in terms of an Hermite expansion,
\begin{equation}
\label{hermite-coll}
{\hat f}^{neq}_i({\bf x}, t) = \frac{w_i}{2T_0} \sum_{n=2}^\infty
\frac{H^{(n)}(\xi_i)}{(n-2)!}{\bf V}({\bf x}, t)^{[n-2]} : {\bf \Pi}^{neq}({\bf x},t)
\end{equation}
Notice the summation starts from $n = 2$ due to the zero mass and momentum
moment values of ${\hat f}^{neq}_i({\bf x}, t)$.
Recall the filter formulation in the previous section, for a lattice
velocity set with an order of isotropy of $2N + 1$, 
in order to ensure a correct (equilibrium plus non-equilibrium) momentum flux, 
the infinite Hermite expansions of (\ref{hermite}) 
and (\ref{hermite-coll}) need to be truncated
to retain $H^{(n)}(\xi_i)$ up to $n \leq 2N - 1$.  As a consequence, the momentum
flux tensor include moments $E^{(n)}$
up to $2N + 1$ supported by the given lattice.

Specifically, for the class of lattice velocity sets of order $2N = 4$,
such as D2Q19, D3Q15 and D3Q19, the truncated non-equilibrium distribution
(\ref{hermite-coll}) becomes
\begin{equation}
\label{filter-1st}
{\hat f}^{neq}_i({\bf x}, t) = \frac{w_i}{2T_0}[(1 + \frac{{\bf c}_i
\cdot {\bf u}({\bf x}, t)}{T_0})(\frac{{\bf c}_i {\bf c}_i}{T_0} - {\bf I})
 - \frac{{\bf c}_i {\bf u}({\bf x},t) + {\bf u}({\bf x},t) {\bf c}_i}{T_0}] :
{\bf \Pi}^{neq}({\bf x},t)
\end{equation}
On the other hand, for the class of lattice velocity sets of order $2N = 6$, 
such as D3Q39, the truncation can be applied at higher order Hermite terms so that
\begin{eqnarray}
\label{filter-2nd}
{\hat f}^{neq}_i({\bf x}, t) = \frac{w_i}{2T_0}[(1 + \frac{{\bf
  c}_i \cdot {\bf u}({\bf x}, t)}{T_0} 
  + \frac{({\bf c}_i \cdot {\bf u}({\bf x}, t))^2}{2T_0^2} - \frac{{\bf u}^2({\bf x},t)}{2T_0})(\frac{{\bf c}_i {\bf c}_i}{T_0} - {\bf I}) \nonumber \\
- (1 + \frac{{\bf c}_i \cdot {\bf u}({\bf x},t)}{T_0})\frac{{\bf c}_i {\bf u}({\bf x}, t) +
{\bf u}({\bf x}, t) {\bf c}_i}{T_0} + \frac{{\bf u}({\bf x},t) {\bf u}({\bf x},t)}{T_0}] :
{\bf \Pi}^{neq}({\bf x},t)
\end{eqnarray}
It is obvious that (\ref{filter-1st}) and (\ref{filter-2nd}) result in
the same momentum flux ${\bf \Pi}^{neq}({\bf x},t)$ as the full non-equilibrium distribution
(\ref{filter-coll}) (or equivalently (\ref{hermite-coll})). 
As a consequence, similar to the basic filter formulation above, the right hand side
of the resulting lattice Boltzmann equation (\ref{lbe}) is given by (\ref{post-1}).
The only difference in the extended filtered collision formulation is 
its non-equilibrium distribution is defined by the
truncated expansion of (\ref{hermite-coll}) (e.g., (\ref{filter-1st}) and (\ref{filter-2nd}))
instead of (\ref{filter-Pi}). 

\section{Discussion}

In this paper, we present an extended filter collision formulation
to enhance Galilean invariance. Indeed, Galilean invariance
is achieved asymptotically as the order of the supporting lattice increases so that the
fully Galilean invariant infinite Hermite expansions of (\ref{hermite}) 
and (\ref{hermite-coll}) can be truncated to higher order terms.
Through the analysis above, we can now
easily reinterpret the previously known collision form of (\ref{filter-Pi}) as the
0th order approximation to the fully Galilean invariant 
non-equilibrium distribution form (\ref{filter-coll}), 
for it only includes terms independent of the local 
fluid velocity ${\bf u}({\bf x},t)$.  The leading order correction 
to (\ref{filter-Pi}) is given by (\ref{filter-1st}) that also includes 
additional terms linearly proportional to ${\bf u}({\bf x},t)$. 
For higher order lattice velocity sets such as D3Q39,
one can go to higher order corrections such as (\ref{filter-2nd})
that include additional terms of ${\bf u}({\bf x},t)$ squared or higher powers.
Although the Galilean invariance is not exactly satisfied
at any finite truncation of the Hermite expansions, the errors are moved 
to higher powers of fluid velocity (or Mach number) as the truncation
is moved to higher order Hermite terms supported by the order of a given lattice.
Therefore, one is able to simulate higher speed fluid flows by an LBM
of higher order Galilean accurate LBM formulated according to
this framework. Indeed, the new filter formulation
has shown to give substantially increased stability and accuracy in simulations
of high speed flows \cite{highspeedpapers}. These are impossible
to achieve with BGK or with the pre-existing filtered collision models \cite{Zhang,Rot,chopard}. 
Furthermore, the new filter collision formulation has also been extended 
for flows with multiple phases and components
showing overall improvements over other known collision models \cite{hiroshi}.
The overall framework of the filter collision procedure has been patented \cite{patent}.
It is also important to notice that same expressions as (\ref{filter-1st}) and 
(\ref{filter-2nd}) have been subsequently produced independently via
an iterative procedure by Malaspinas et all \cite{malaspinas}.
There are, however, some critical differences in the two theoretical approaches.  
First of all, our formulation presented here has a clear physical picture of achieving
Galilean invariance via particle relative velocity.
Secondly, our formulation admits fully Galilean invariant compact
forms for both the equilibrium and non-equilibrium distribution functions,
and the two compact forms have exact and explicit corresponding infinite
Hermite expansions.  Lastly, a filter collision operator can be directly obtained 
by simply applying a proper truncation to the infinite expansions according to 
the order of the given supporting lattice velocity set, instead of
going through a convoluted iteration procedure. We also want to mention a parallel approach which aims to recover Galilean invariance in collision operators via the so called “cascaded” multiple-relaxation time (MRT) LBM formulation \cite{Geier,Asinari}.

In the above, we present in detail the extension of the filter collision formulation
involving only the non-equilibrium momentum flux tensor.
However, the overall procedure is more general in that it can include any other non-equilibrium
moments in the resulting non-equilibrium distribution function.  
Indeed, it is straightforward to also include the
energy flux tensor ${\bf Q}^{neq}({\bf x},t)$, similar to that in basic filter collision form
(\ref{filter2}).  Explicitly, for a generic fully Galilean invariant form (\ref{filter-coll}),
we can write
\begin{equation}
\label{pq} 
{\hat f}^{neq}_i({\bf x}, t) \equiv {\hat f}^{neq,p}_i({\bf x}, t)
+ \theta {\hat f}^{neq,q}_i({\bf x}, t)
\end{equation}
where parameter $\theta$ is introduced for fluids
with non-unity Prandtl numbers.  
${\hat f}^{neq,p}_i({\bf x}, t)$ is given by (\ref{filter-coll}) in the above.
On the other hand, using the same argument in terms of particle relative velocity
${\bf c}'_i({\bf x},t)$, we can write ${\hat f}^{neq,q}_i({\bf x}, t)$ as follows,
\begin{equation}
\label{filter-Q}
{\hat f}^{neq,q}_i({\bf x}, t) = 
\frac{f^{eq}_i({\bf x},t)}{6T_0^3} [{\bf c}'_i({\bf x},t) {\bf c}'_i({\bf
    x},t) {\bf c}'_i({\bf x}, t) - 3 {\bf c}'_i({\bf x}, t) T_0 {\bf I}] \vdots
{\tilde {\bf Q}}^{neq}({\bf x}, t)
\end{equation}
where the energy flux tensor ${\tilde {\bf Q}}^{neq}({\bf x}, t)$ is defined in (\ref{Q_def}) 
in terms of the relative particle velocity.  It is easily seen that 
${\tilde {\bf Q}}^{neq}({\bf x}, t)$ is a simple linear combination of 
the absolute energy flux tensor ${\bf Q}^{neq}({\bf x},t)$ 
in (\ref{coeff-3}) and a direct tensor product of the non-equilibrium momentum flux 
tensor ${\bf \Pi}^{neq}({\bf x},t)$ and the fluid velocity ${\bf u}({\bf x},t)$ as well as
their symmetric permutations.  It has been shown that the above LBM
formulation results in the correct hydrodynamic equation for non-unity Prandtl 
numbers, while all the pre-existing LBM models have shown to suffer
strong velocity dependent artifacts \cite{Prandtl}.

\vspace{0.3in}

{\bf Acknowledgment:} We are grateful to Ilya Staroselsky, Hongli Fan and Hiroshi Otomo for insightful discussions.  

We thank the 10th Anniversary Program “Turbulent Mixing and Beyond”-17 for 
inviting one of us (H. C.) and we value the effort this program makes 
organizing the scientific community to search for unifying conceptual themes 
between description of non-equilibrium processes of various phenomenology. 
We believe that looking for unification becomes more and more important inasmuch 
as research areas keep specializing and branching. We also believe that in 
addition to the conceptual value of the unified understanding of non-equilibrium 
dynamics that is being sought, it should bring concrete advances in science 
and technology as has been historically the case for similar undertakings. 
To that end, we identify with the TMB program’s effort to connect fundamental 
physics of non-equilibrium processes with practical applications and we hope 
that a small contribution presented here may support this view.

\appendix{Derivation of the Hermite expanded form}

Here we show in detail mathematically that the compact form 
in (\ref{filter-coll}) for the non-equilibrium
distribution function is exactly equal to the infinite 
Hermite expansion in (\ref{hermite-coll}).

Rewrite for convenience the expression (\ref{filter-coll}) below,
\begin{equation}
\label{filter-coll-A}
{\hat f}^{neq}_i = \frac{f^{eq}_i}{2 \rho T_0} 
[\frac{{\bf c}'_i {\bf c}'_i}{T_0} - {\bf I}] : {\bf \Pi}^{neq}
\end{equation}
where $\xi_i = {\bf c}_i/T_0^{1/2}$, 
and ${\bf c}'_i = {\bf c}_i - {\bf u}$ is the particle relative velocity.  
The equilibrium distribution is given 
in terms of an Hermite expansion in (\ref{hermite}), 
\begin{equation}
\label{hermite-A}
f^{eq}_i = w_i \rho \sum_{n=0}^{\infty} \frac{H^{(n)}(\xi_i)}{n!} {\bf V}^{[n]}
\end{equation}
Combining (\ref{filter-coll-A}) and (\ref{hermite-A}) together with the definitions
of $\xi_i$ and ${\bf c}'_i$, we have
\begin{eqnarray}
\label{expan-A}
{\hat f}^{neq}_i &=& \frac{w_i}{2 T_0} \sum_{n=0}^\infty \frac{H^{(n)}(\xi_i)}{n!} {\bf V}^{[n]}
[\frac{{\bf c}'_i {\bf c}'_i}{T_0} - {\bf I}] : {\bf \Pi}^{neq}
\nonumber \\
&=& \frac{w_i}{2 T_0} \sum_{n=0}^{\infty} \frac{H^{(n)}(\xi_i)}{n!} {\bf V}^{[n]}
[(\xi_i\xi_i - {\bf I}) - ({\bf V}\xi_i + \xi_i {\bf V}) + {\bf V}{\bf V}]:{\bf \Pi}^{neq}
\end{eqnarray}
The above expression can be rewritten in terms of Cartesian component form,
\begin{eqnarray}
\label{comp-A}
{\hat f}^{neq}_i &=& \frac{w_i}{2 T_0} \sum_{n=0}^\infty 
\frac{H^{(n)}_{\alpha_1\ldots \alpha_n}}{n!} V_{\alpha_1}\cdots V_{\alpha_n}
[(\xi_{i\beta}\xi_{i\gamma} - \delta_{\beta\gamma}) \nonumber \\
&-& (V_\beta\xi_{i\gamma} + \xi_{i\beta}V_\gamma) + V_\beta V_\gamma]\Pi^{neq}_{\beta\gamma}
\end{eqnarray}
where sub-indexes $\alpha_1\ldots \alpha_n$, $\beta$ and $\gamma$ are Cartesian components,
and repeated indexes are summed over.
Re-arranging terms with equal power of $V$, expression (\ref{comp-A}) becomes
\begin{eqnarray}
\label{Power-A}
{\hat f}^{neq}_i &=& \frac{w_i}{2 T_0} 
\{ \frac{H^{(0)}} {0!} (\xi_{i\beta}\xi_{i\gamma} - \delta_{\beta\gamma})
\nonumber \\
&+& [ \frac{H^{(1)}_\alpha} {1!} V_\alpha (\xi_{i\beta}\xi_{i\gamma} - \delta_{\beta\gamma})
- \frac{H^{(0)}} {0!} (V_\beta\xi_{i\gamma} + \xi_{i\beta}V_\gamma) ]
\nonumber \\
&+& \sum_{n=1}^\infty [ \frac{H^{(n+1)}_{\alpha_1\ldots \alpha_{n+1}}}{(n+1)!}
V_{\alpha_1}\cdots V_{\alpha_{n+1}} (\xi_{i\beta}\xi_{i\gamma} - \delta_{\beta\gamma})
\nonumber \\
& & - \frac{H^{(n)}_{\alpha_1\ldots \alpha_n}}{n!} V_{\alpha_1}\cdots V_{\alpha_n}
(V_\beta\xi_{i\gamma} + \xi_{i\beta}V_\gamma)
\nonumber \\
& & + \frac{H^{(n-1)}_{\alpha_1\ldots \alpha_{n-1}}}{(n-1)!} V_{\alpha_1}\cdots V_{\alpha_{n-1}}
V_\beta V_\gamma ] \} \Pi^{neq}_{\beta\gamma}
\end{eqnarray}
Now let us examine term by term in (\ref{Power-A}).
For $O(V^{0})$,
\begin{equation}
\label{Order0}
\frac{H^{(0)}} {0!} (\xi_{i\beta}\xi_{i\gamma} - \delta_{\beta\gamma})
= (\xi_{i\beta}\xi_{i\gamma} - \delta_{\beta\gamma}) = H^{(2)}_{\beta\gamma}
\end{equation}
For $O(V^{1})$,
\begin{eqnarray}
\label{Order1}
& & \frac{H^{(1)}_\alpha} {1!} V_\alpha (\xi_{i\beta}\xi_{i\gamma} - \delta_{\beta\gamma})
- \frac{H^{(0)}} {0!} (V_\beta\xi_{i\gamma} + \xi_{i\beta}V_\gamma)
\nonumber \\
& & = \xi_\alpha V_\alpha (\xi_{i\beta}\xi_{i\gamma} - \delta_{\beta\gamma})
- (V_\beta\xi_{i\gamma} + \xi_{i\beta}V_\gamma)
\nonumber \\
& & = (\xi_\alpha\xi_\beta\xi_\gamma - \xi_\alpha\delta_{\beta\gamma}
- \xi_\beta\delta_{\gamma\alpha} - \xi_\gamma\delta_{\alpha\beta} )V_\alpha
\nonumber \\
& & = H^{(3)}_{\alpha\beta\gamma} V_\alpha
\end{eqnarray}
Substituting (\ref{Order0}) and (\ref{Order1}) into (\ref{Power-A}), it becomes
\begin{eqnarray}
\label{Order-A}
{\hat f}^{neq}_i &=& \frac{w_i}{2 T_0} 
\{ H^{(2)}_{\beta\gamma} + H^{(3)}_{\alpha\beta\gamma} V_\alpha
\nonumber \\
&+& \sum_{n=1}^\infty [ \frac{H^{(n+1)}_{\alpha_1\ldots \alpha_{n+1}}}{(n+1)!}
V_{\alpha_1}\cdots V_{\alpha_{n+1}} (\xi_{i\beta}\xi_{i\gamma} - \delta_{\beta\gamma})
\nonumber \\
& & - \frac{H^{(n)}_{\alpha_1\ldots \alpha_n}}{n!} V_{\alpha_1}\cdots V_{\alpha_n}
(V_\beta\xi_{i\gamma} + \xi_{i\beta}V_\gamma)
\nonumber \\
& & + \frac{H^{(n-1)}_{\alpha_1\ldots \alpha_{n-1}}}{(n-1)!} V_{\alpha_1}\cdots V_{\alpha_{n-1}}
V_\beta V_\gamma ] \} \Pi^{neq}_{\beta\gamma}
\end{eqnarray}
The remaining task is to deal with terms involving $O(V^{n+1})$ ($n \geq 1$).
It is straightforward to recognize that the expression inside the summation 
$\sum_{n=1}^{\infty}$ can be recast below by extracting all the ``$V$''s,
\begin{eqnarray}
\label{sum-terms}
&S& \equiv [\frac{H^{(n+1)}_{\alpha_1\ldots \alpha_{n+1}}}{(n+1)!}
(\xi_{i\beta}\xi_{i\gamma} - \delta_{\beta\gamma})
\nonumber \\
& & - \frac{H^{(n)}_{\alpha_1\ldots \alpha_n}}{n!} 
(\delta_{\alpha_{n+1}\beta}\xi_{i\gamma} + \delta_{\alpha_{n+1}\gamma}\xi_{i\beta})
\nonumber \\
& & + \frac{H^{(n-1)}_{\alpha_1\ldots \alpha_{n-1}}}{(n-1)!} 
\delta_{\alpha_n\beta}\delta_{\alpha_{n+1}\gamma} ] V_{\alpha_1}\cdots V_{\alpha_{n+1}}
\end{eqnarray}
Therefore, the task is to prove that the expression in (\ref{sum-terms}) is equal to
\[ 
\frac {H^{(n+3)}_{\alpha_1\ldots \alpha_{n+1}\beta\gamma}}{(n+1)!}
V_{\alpha_1}\cdots V_{\alpha_{n+1}}
\]
For this purpose, we utilize the recursive relation of Hermite polynomials,
\begin{equation}
\label{recurG}
H^{(n+1)}_{\alpha_1\ldots \alpha_{n+1}} = \xi_{i\alpha_{n+1}} H^{(n)}_{\alpha_1\ldots \alpha_n}
- \sum_{k=1}^n \delta_{\alpha_{n+1}\alpha_k} 
H^{(n-1)}_{\alpha_1\ldots\alpha_{k-1}\alpha_{k+1}\ldots\alpha_n}
\end{equation}
Applying the recursive relation repeatedly, through straightforward algebra we have
\begin{eqnarray}
\label{recurN}
& & H^{(n+3)}_{\alpha_1\ldots \alpha_{n+1}\beta\gamma} 
\nonumber \\
& & = \xi_{i\gamma} H^{(n+2)}_{\alpha_1\ldots \alpha_{n+1}\beta}
- \sum_{m=1}^{n+2} \delta_{\gamma\alpha_m} 
H^{(n+1)}_{\alpha_1\ldots\alpha_{m-1}\alpha_{m+1}\ldots\alpha_{n+1}\beta}
\nonumber \\
& & = \xi_{i\gamma} [\xi_{i\beta} H^{(n+1)}_{\alpha_1\ldots \alpha_{n+1}}
- \sum_{l=1}^{n+1}\delta_{\beta\alpha_l} 
H^{(n)}_{\alpha_1\ldots\alpha_{l-1}\alpha_{l+1}\ldots\alpha_{n+1}} ]
\nonumber \\
& & - \delta_{\beta\gamma} H^{(n+1)}_{\alpha_1\ldots \alpha_{n+1}}
- \sum_{m=1}^{n+1}\delta_{\gamma\alpha_m} 
H^{(n+1)}_{\alpha_1\ldots\alpha_{m-1}\alpha_{m+1}\ldots\alpha_{n+1}\beta}
\nonumber \\
& & = H^{(n+1)}_{\alpha_1\ldots \alpha_{n+1}} (\xi_{i\beta}\xi_{i\gamma} - \delta_{\beta\gamma})
\nonumber \\
& & - \xi_{i\gamma} \sum_{l=1}^{n+1}\delta_{\beta\alpha_l} 
H^{(n)}_{\alpha_1\ldots\alpha_{l-1}\alpha_{l+1}\ldots\alpha_{n+1}}
\nonumber \\
& & - \sum_{m=1}^{n+1}\delta_{\gamma\alpha_m} 
H^{(n+1)}_{\alpha_1\ldots\alpha_{m-1}\alpha_{m+1}\ldots\alpha_{n+1}\beta}
\nonumber \\
& & = H^{(n+1)}_{\alpha_1\ldots \alpha_{n+1}} (\xi_{i\beta}\xi_{i\gamma} - \delta_{\beta\gamma})
\nonumber \\
& & - \xi_{i\gamma} \sum_{l=1}^{n+1}\delta_{\beta\alpha_l} 
H^{(n)}_{\alpha_1\ldots\alpha_{l-1}\alpha_{l+1}\ldots\alpha_{n+1}}
\nonumber \\
& & - \sum_{m=1}^{n+1}\delta_{\gamma\alpha_m} 
[ \xi_{i\beta}H^{(n)}_{\alpha_1\ldots\alpha_{m-1}\alpha_{m+1}\ldots\alpha_{n+1}} 
\nonumber \\
& & - \sum_{k=1, k\neq m}^n \delta_{\beta\alpha_k}
H^{(n-1)}_{\alpha_1\ldots\alpha_{k-1}\alpha_{k+1}\ldots\alpha_{n+1}} ]
\end{eqnarray}
With some slight rearrangement of terms in (\ref{recurN}), we get
\begin{eqnarray}
\label{recurR}
& & H^{(n+3)}_{\alpha_1\ldots \alpha_{n+1}\beta\gamma} 
\nonumber \\
& & = H^{(n+1)}_{\alpha_1\ldots \alpha_{n+1}} (\xi_{i\beta}\xi_{i\gamma} - \delta_{\beta\gamma})
\nonumber \\
& & - \sum_{l=1}^{n+1} 
H^{(n)}_{\alpha_1\ldots\alpha_{l-1}\alpha_{l+1}\ldots\alpha_{n+1}}
(\xi_{i\beta}\delta_{\gamma\alpha_l} + \xi_{i\gamma}\delta_{\beta\alpha_l})
\nonumber \\
& & + \sum_{k=1, k\neq m}^n \sum_{m=1}^{n+1}\delta_{\gamma\alpha_m}\delta_{\beta\alpha_k}
H^{(n-1)}_{\alpha_1\ldots\alpha_{k-1}\alpha_{k+1}\ldots\alpha_{n+1}}
\end{eqnarray}
Consequently, when (\ref{recurR}) is multiplied with symmetric direct vector
product of ${\bf V}^{[n+1]}$ (i.e., $V_{\alpha_1}\cdots V_{\alpha_{n+1}}$), we obtain
\begin{eqnarray}
\label{Prod-A}
& & H^{(n+3)}_{\alpha_1\ldots \alpha_{n+1}\beta\gamma} V_{\alpha_1}\cdots V_{\alpha_{n+1}}
\nonumber \\
& & = [ H^{(n+1)}_{\alpha_1\ldots \alpha_{n+1}} (\xi_{i\beta}\xi_{i\gamma} - \delta_{\beta\gamma}) \nonumber \\
& & - (n + 1) H^{(n)}_{\alpha_1\ldots\alpha_n}
(\xi_{i\beta}\delta_{\gamma\alpha_{n+1}} + \xi_{i\gamma}\delta_{\beta\alpha_{n+1}})
\nonumber \\
& & + (n + 1)n H^{(n-1)}_{\alpha_1\ldots \alpha_{n-1}} 
\delta_{\beta\alpha_n}\delta_{\gamma\alpha_{n+1}} ] V_{\alpha_1}\cdots V_{\alpha_{n+1}}
\end{eqnarray}
Comparing (\ref{Prod-A}) and (\ref{sum-terms}), we immediately see that
\begin{equation}
\label{S-term}
S = \frac{H^{(n+3)}_{\alpha_1\ldots \alpha_{n+1}\beta\gamma}}{(n+1)!} 
V_{\alpha_1}\cdots V_{\alpha_{n+1}}
\end{equation}
Substituting (\ref{S-term}) above into (\ref{Order-A}), we finally prove the
following,
\begin{eqnarray}
\label{Noneq-A}
{\hat f}^{neq}_i &=& \frac{w_i}{2 T_0} 
\{ H^{(2)}_{\beta\gamma} + H^{(3)}_{\alpha\beta\gamma} V_\alpha
\nonumber \\
&+& \sum_{n=1}^\infty \frac{H^{(n+3)}_{\alpha_1\ldots \alpha_{n+1}\beta\gamma}}{(n+1)!}
V_{\alpha_1}\cdots V_{\alpha_{n+1}} \} \Pi^{neq}_{\beta\gamma}
\end{eqnarray}
Or equivalently in vector product notation,
\begin{equation}
\label{NoneqV-A}
{\hat f}^{neq}_i = \frac{w_i}{2 T_0} 
\sum_{n=2}^\infty \frac{H^{(n)}}{(n-2)!}{\bf V}^{[n-2]}: {\bf \Pi}^{neq}
\end{equation}
This is exactly the same form as in (\ref{hermite-coll}), hence the proof is complete.

\end{document}